\documentstyle[floats,prl,aps]{revtex}
\begin{document}
\renewcommand{\thefootnote}{\fnsymbol{footnote}}
\draft
\title{\Large\bf 
Nine classes of integrable boundary conditions for the eight-state
supersymmetric fermion model}

\author{ Yao-Zhong Zhang \footnote {yzz@maths.uq.edu.au}
             and 
        Huan-Qiang Zhou \footnote{hqz@maths.uq.edu.au}}

\address{Department of Mathematics,University of Queensland,
		     Brisbane, Qld 4072, Australia}

\maketitle

\vspace{10pt}

\begin{abstract}
Nine classes of integrable boundary conditions for the eight-state 
supersymmetric model of strongly correlated fermions
are presented. The boundary systems are solved 
by using the coordinate Bethe ansatz method and the Bethe ansatz
equations for all nine cases are given. 
\end{abstract}




\def\a{\alpha}
\def\b{\beta}
\def\d{\delta}
\def\e{\epsilon}
\def\g{\gamma}
\def\k{\kappa}
\def\l{\lambda}
\def\o{\omega}
\def\t{\theta}
\def\s{\sigma}
\def\D{\Delta}
\def\L{\Lambda}


\def\beq{\begin{equation}}
\def\eeq{\end{equation}}
\def\bea{\begin{eqnarray}}
\def\eea{\end{eqnarray}}
\def\ba{\begin{array}}
\def\ea{\end{array}}
\def\no{\nonumber}
\def\le{\langle}
\def\re{\rangle}
\def\lt{\left}
\def\rt{\right}

\newcommand{\sect}[1]{\setcounter{equation}{0}\section{#1}}
\renewcommand{\theequation}{\thesection.\arabic{equation}}
\newcommand{\reff}[1]{eq.~(\ref{#1})}

\vskip.3in

\sect{Introduction}

Lattice integrable models with open-boundary conditions are one of the recent
developments which deserves careful elaborations. 
As many systems in nature are confined in finite boxes (or intervals for
one-dimensional systems), the effects
of boundaries are very significant.  This is
particularly so for integrable systems since boundary conditions generally
spoil the integrability of the bulk models. Therefore,
the problem of how to extend a bulk integrable model to include integrable
boundary conditions becomes very important.

A systematic method for treating integrable lattice models with boundaries,
that is the boundary quantum inverse scattering method (QISM),
has been developed by Sklyanin \cite{Skl88} and generalized in 
\cite{Mez91,Bra97}. Within this framework, the integrable boundary
conditions are determined by boundary K-matrices obeying the
(graded) reflection equations. 

Integrable correlated fermion systems constitute an important class
of lattice integrable models, which have recently attracted much
attention \cite{Ess94,Bar91,Bra95,Bar95,Gou97}. In \cite{Gou97}, we
proposed two new integrable models with Lie superalgebra $gl(3|1)$ and
quantum superalgebra $U_q[gl(3|1)]$ symmetries, respectively. These
are eight-state fermion models with correlated single-particle and
pair hoppings, uncorrelated triple-particle hopping and two- and
three-particle on-site interactions. By eight-state, we mean that
at a given lattice site $j$ of the length $L$
there are eight possible states:
\bea
&&|0\re\,,~~~c_{j,+}^\dagger|0\re\,,~~~
  c_{j,0}^\dagger|0\re\,,~~~ c_{j,-}^\dagger|0\re,\no\\
&&c_{j,+}^\dagger c_{j,0}^\dagger|0\re\,,~~~ 
  c_{j,+}^\dagger c_{j,-}^\dagger|0\re\,,~~~ 
  c_{j,0}^\dagger c_{j,-}^\dagger|0\re\,,~~~ 
  c_{j,+}^\dagger c_{j,0}^\dagger c_{j,-}^\dagger|0\re\,,\label{states}
\eea
where $c_{j,\a}^\dagger$ ($c_{j,\a}$) denotes a fermionic creation
(annihilation) operator which creates (annihilates) a fermion
of species $\a=+,\;0,\;-$ at
site $j$; these operators satisfy the anti-commutation relations given by
$\{c_{i,\a}^\dagger, c_{j,\b}\}=\d_{ij}\d_{\a\b}$.

In a series of papers, we have constructed
a large number of integrable boundary conditions for various models
of strongly correlated electrons \cite{Zha97,Ge97,Bra97}.
In this paper, we are
concerned with the integrable eight-state fermion model with Lie
superalgebra $gl(3|1)$ symmetry. We present nine classes of boundary
conditions for this model, all of which are shown to be integrable
by the graded boundary QISM recently formulated in \cite{Bra97}.
We solve the boundary systems by using the coordinate Bethe
ansatz method and derive the Bethe ansatz equations for all nine cases.

This paper is organized as follows. In Section II the boundary model
Hamiltonians are described. In the following sections we establish the
quantum integrability for all these boundary conditions,and derive the
corresponding Bethe ansatz equations in terms of the coordinate Bethe
ansatz method. The last section is devoted to the conclusion.

\sect{Boundary Model Hamiltonians}

We consider the following Hamiltonian with boundary terms
\beq
H=\sum _{j=1}^{L-1} H_{j,j+1}^{\rm bulk} + H^{\rm boundary}_{lt} 
+H^{\rm boundary}_{rt},\label{h}
\eeq
where $H^{\rm boundary}_{lt},~H^{\rm boundary}_{rt}$ are 
respectively left and right
boundary terms whose explicit forms are given below, and
$H^{\rm bulk}_{j,j+1}$ is the Hamiltonian density of the eight-state 
supersymmetric $U$ model \cite{Gou97}
\bea
H^{\rm bulk}_{j,j+1}(g)&=&-\sum_\a(c_{j,\a}^\dagger c_{j+1,\a}+{\rm h.c.})
  \;\exp\lt\{-\frac{\eta}{2} \sum_{\b(\neq\a)}(n_{j,\b}+n_{j+1,\b})
  +\frac{\zeta}{2}
  \sum_{\b\neq\g(\neq\a)}(n_{j,\b}n_{j,\g}
  +n_{j+1,\b} n_{j+1,\g})\rt\}\no\\
& &-\frac{1}{2(g+1)}\sum_{\a\neq\b\neq\g}(c_{j,\a}^\dagger c_{j,\b}^\dagger 
  c_{j+1,\b}c_{j+1,\a}+{\rm h.c.})
  \;\exp\lt\{-\frac{\xi}{2}  (n_{j,\g}+n_{j+1,\g})\rt\}\no\\
& &-\frac{2}{(g+1)(g+2)}\lt(c_{j,+}^\dagger c_{j,0}^\dagger 
  c_{j,-}^\dagger c_{j+1,-} c_{j+1,0} c_{j+1,+}+{\rm h.c.}\rt)\no\\
& & +\sum_\a (n_{j,\a}+n_{j+1,\a})-\frac{1}{2(g+1)}\sum_{\a\neq\b}
  (n_{j,\a}n_{j,\b}+n_{j+1,\a}n_{j+1,\b})\no\\
& &+\frac{2}{(g+1)(g+2)}(n_{j,+}n_{j,0}n_{j,-}+n_{j+1,+}n_{j+1,0}
  n_{j+1,-}),\label{hamiltonian}
\eea
where $n_{j\a}$ is the number density operator
$n_{j\a}=c_{j\a}^{\dagger}c_{j\a}$,
$n_j=n_{j+}+n_{j0}+n_{j-}$; and
\beq
\eta=-\ln\frac{g}{g+1},~~~\zeta=\ln(g+1)-\frac{1}{2}
  \ln g(g+2),~~~\xi=-\ln\frac{g}{g+2}.
\eeq
We claim that the boundary Hamiltonain (\ref{h}) is integrable under
the boundary conditions:
\bea
{\rm Case ~(i)}:~~H^{\rm boundary}_{lt}&=&-\frac {2g}
  {2-\xi^I_-}\lt(n_1-
  \frac {2}{\xi^I_-}
 ( n_{1+}n_{10}+n_{10}n_{1-}+n_{1+}n_{1-})
  +\frac {8}{\xi^I _- (2+\xi^I _-)}n_{1+} n_{10} n_{1-} \rt),\no\\
H^{\rm boundary}_{rt}&=&
  -\frac {2g}{2-\xi^I_+}\lt(n_L-
  \frac {2}{\xi^I_+}
 ( n_{L+}n_{L0}+n_{L0}n_{L-}+n_{L+}n_{L-})
  +\frac {8}{\xi^I _+ (2+\xi^I _+)}n_{L+} n_{L0} n_{L-} \rt)
  ;\label{boundary11}\\
{\rm Case ~(ii)}:~~H^{\rm boundary}_{lt}&=&\frac {2g}
  {\xi^{II}_-}\lt(n_{10}+n_{1-}- \frac{2}{2-\xi^{II}_-}n_{10}n_{1-}\rt),\no\\
H^{\rm boundary}_{rt}&=& \frac{2g}{\xi^{II}_+}\lt(n_{L0}+n_{L-}-
  \frac {2}{2-\xi^{II}_+}n_{L0}n_{L-}\rt);\label{boundary22}\\
{\rm Case ~(iii)}:~~H^{\rm boundary}_{lt}&=&\frac {2g}{\xi^{III}_-}\;
   n_{1-},~~~~~~
H^{\rm boundary}_{rt}= \frac{2g}{\xi^{III}_+}\;n_{L-},\label{boundary33}
\eea
plus six others, numbered from Case (iv) to Case (ix) below,
which are built from the above three cases by using the fact that
boundary conditions on the left and on the right end of an open lattice
chain can be chosen independently. 
Throughout, $\xi^{a}_{\pm}(a=I,II,III)$ are some parameters describing the
boundary effects. 

\sect{Quantum Integrability}

Quantum integrability of the boundary conditions proposed in the
previous section can be established by means of 
the (graded) boundary QISM recently formulated in \cite{Bra97}. Indeed, 
the integrability corresponding to the above Case (i) has been shown in
\cite{Ge97}. We now establish the integrability for the remaining eight
cases. We first search for boundary K-matrices which
satisfy the graded reflection equations:
\bea
&&R_{12}(u_1-u_2)\stackrel {1}{K}_-(u_1) R_{21}(u_1+u_2)
  \stackrel {2}{K}_-(u_2)
=  \stackrel {2}{K}_-(u_2) R_{12}(u_1+u_2)
  \stackrel {1}{K}_-(u_1) R_{21}(u_1-u_2),  \label{reflection1}\\
&&R_{21}^{st_1 ist_2}(-u_1+u_2)\stackrel {1}{K_+^{st_1}}
  (u_1) R_{12}(-u_1-u_2+4)
  \stackrel {2}{K_+^{ist_2}}(u_2)\no\\
&&~~~~~~~~~~~~~~~~~~~~~~~~~=\stackrel {2}{K_+^{ist_2}}(u_2) R_{21}(-u_1-u_2+4)
  \stackrel {1}{K_+^{st_1}}(u_1) R_{12}^{st_1 ist_2}(-u_1+u_2)
  ,\label{reflection2}
\eea
where $R(u)\in End(V\otimes V)$, with $V$ being 8-dimensional
representation of $gl(3|1)$, is the R-matrix of the eight-state 
supersymmetric $U$ model \cite{Gou97}, and $R_{21}(u)=P_{12}R_{12}(u)P_{12}$
with $P$ being the graded permutation operator;
the supertransposition $st_{\mu}~(\mu =1,2)$ 
is only carried out in the
$\mu$-th factor superspace of $V \otimes V$, whereas $ist_{\mu}$ denotes
the inverse operation of  $st_{\mu}$. 

It can be checked that there are three different diagonal boundary 
K-matrices \footnote{Non-diagonal K-matrices exist but they are beyond
the scope of this paper.}, 
$K^I_-(u),~ K^{II}_-(u),~K^{III}_-(u)$, which solve the
first reflection equation (\ref{reflection1}):
\bea
K^I_-(u)&=&\frac {1}{ \xi^I_-(2-\xi^I_-)(2+\xi^I_-)} {\rm diag}\lt ( 
  A^I_-(u),B^I_-(u),B^I_-(u),B^I_-(u),C^I_-(u),C^I_-(u),
  C^I_-(u),D^I_-(u)\rt),\no\\
K^{II}_-(u)&=&\frac {1}{ \xi^{II}_-(2-\xi^{II}_-)} {\rm diag}\lt ( 
  A^{II}_-(u),A^{II}_-(u),B^{II}_-(u),B^{II}_-(u),B^{II}_-(u),B^{II}_-(u),
  C^{II}_-(u),C^{II}_-(u)\rt),\no\\
K^{III}_-(u)&=&\frac {1}{ \xi^{III}_-} {\rm diag}\lt ( 
  A^{III}_-(u),A^{III}_-(u),A^{III}_-(u),B^{III}_-(u),
  A^{III}_-(u),B^{III}_-(u),
  B^{III}_-(u),B^{III}_-(u)\rt),
\eea
where
\bea
A^I_-(u)&=&(-\xi^I_-+u)(2-\xi^I_-+u)(-2-\xi^I_- +u),\no\\
B^I_-(u)&=&(-\xi^I_-+u)(2-\xi^I_--u)(-2-\xi^I_- +u),\no\\
C^I_-(u)&=&(-\xi^I_--u)(2-\xi^I_--u)(-2-\xi^I_-+u),\no\\
D^I_-(u)&=&(-\xi^I_--u)(2-\xi^I_--u)(-2-\xi^I_--u),\no\\
A^{II}_-(u)&=&(\xi^{II}_-+u)(2-\xi^{II}_--u),\no\\
B^{II}_-(u)&=&(\xi^{II}_--u)(2-\xi^{II}_--u),\no\\
C^{II}_-(u)&=&(\xi^{II}_--u)(2-\xi^{II}_-+u),\no\\
A^{III}_-(u)&=&(\xi^{III}_-+u),~~~~~~
  B^{III}_-(u)=(\xi^{III}_--u).
\eea
The corresponding K-matrices, $K^I_+(u),~K^{II}_+(u),~K^{III}_+(u)$,
can be obtained from the isomorphism of the
two reflection equations. Indeed, given a solution
$K^a_- (u)$ of (\ref{reflection1}), then $K^a_+(u)$ defined by
\beq
{K^a_+}^{st}(u) =  K^a_-(-u+2),~~~~~a=I,\,II,\;III,\label{t+t-}
\eeq
are solutions of (\ref{reflection2}). 
The proof follows from some algebraic computations upon
substituting (\ref{t+t-}) into  
(\ref{reflection2}) and making use
of the properties of the R-matrix .

Following  Sklyanin's arguments \cite{Skl88}, one
may show that the quantity ${\cal T}_-(u)$ given by
\beq
{\cal T}_-(u) = T(u) {K}_-(u) T^{-1}(-u),~~~~~~ 
T(u) = R_{0L}(u)\cdots R_{01}(u),
\eeq
satisfies the same relation as $K_-(u)$:
\beq
R_{12}(u_1-u_2)\stackrel {1}{\cal T}_-(u_1) R_{21}(u_1+u_2)
  \stackrel {2}{\cal T}_-(u_2)
=  \stackrel {2}{\cal T}_-(u_2) R_{12}(u_1+u_2)
  \stackrel {1}{\cal T}_-(u_1) R_{21}(u_1-u_2).
\eeq
Thus if one defines the boundary transfer matrix $\tau(u)$ as
\beq
\tau(u) = str (K_+(u){\cal T}_-(u))=str\lt(K_+(u)T(u)K_-(u)T^{-1}(-u)\rt),
\eeq
then it can be shown \cite{Bra97} that
$[\tau(u_1),\tau(u_2)] = 0$. Since $K_\pm(u)$ can be taken as 
$K^I_\pm(u),~K^{II}_\pm(u)$ and $K^{III}_\pm(u)$, respectively, we have nine
possible choices of boundary transfer matrices:
\beq
\tau^{(a,b)}(u)=str\lt(K^a_+(u)T(u)K^b_-(u)T^{-1}(-u)\rt),~~~~
  a,\;b=I,\;II,\;III,\label{t-matrices}
\eeq
which reflects the fact that the boundary conditions on the left end
and on the right end of the open lattice chain are independent. 

Now it can be shown  that 
Hamiltonians corresponding to all nine boundary conditions are related
to the second derivative of
the boundary transfer matrix
$\tau^{(a,b)} (u)$ (up to an unimportant additive constant)
\bea
H&=&2g \; H^{(a,b)},\no\\
H^{(a,b)}&=&\frac {\tau^{(a,b)''} (0)}{4(V+2W)}=
  \sum _{j=1}^{L-1} H^R_{j,j+1} + \frac {1}{2} \stackrel {1}{K^{b'}}_-(0)
+\frac {1}{2(V+2W)}\lt[str_0\lt(\stackrel {0}{K^a}_+(0)G_{L0}\rt)\rt.\no\\
& &\lt.+2\,str_0\lt(\stackrel {0}{K^{a'}}_+(0)H_{L0}^R\rt)+
  str_0\lt(\stackrel {0}{K^a}_+(0)\lt(H^R_{L0}\rt)^2\rt)\rt],\label{derived-h}
\eea
where 
\bea
V&=&str_0 K^{a'}_+(0),
~~W=str_0 \lt(\stackrel {0}{K^a}_+(0) H_{L0}^R\rt),\no\\
H^R_{j,j+1}&=&P_{j,j+1}R'_{j,j+1}(0),
~~G_{j,j+1}=P_{j,j+1}R''_{j,j+1}(0)
\eea
with $P_{j,j+1}$ denoting the graded permutation operator acting on the $j$-th
and $j+1$-th quantum spaces. More precisely,  Case (i), Case (ii) and
Case (iii) correspond to $H^{(I,I)},~H^{(II,II)}$ and
$H^{(III,III)}$, respectively. We arrange the remaing six cases in
the following order so that we have the correspondence: Case (iv) 
$\leftrightarrow H^{(I,II)}$, Case (v) $\leftrightarrow H^{(II,I)}$,
Case (vi) $\leftrightarrow H^{(I,III)}$, Case (vii)
$\leftrightarrow H^{(III,I)}$, Case (viii) $\leftrightarrow H^{(II,III)}$
and Case (ix) $\leftrightarrow H^{(III,II)}$.

Let us also remark that for general boundary parameters $\xi_\pm^a$ 
the boundary terms listed above break
the original $gl(3|1)$ symmetry of the bulk model into $U(1)\times
U(1)\times U(1)$ symmetry (generated by fermion number operators).

\sect{Bethe Ansatz Solutions}

Having established the quantum integrability of the boundary model,
we now  solve it by using
the coordinate space Bethe ansatz method.
Following \cite{Asa96,Zha97,Bra97,Ge97},
we assume that the eigenfunction of Hamiltonian (\ref{hamiltonian}) 
takes the form
\bea
| \Psi \rangle& =&\sum _{\{(x_j,\a_j)\}}\Psi _{\a_1,\cdots,\a_N}
  (x_1,\cdots,x_N)c^\dagger
  _{x_1\a_1}\cdots c^\dagger_{x_N\a_N} | 0 \rangle,\no\\
\Psi_{\a_1,\cdots,\a_ N}(x_1,\cdots,x_N)
&=&\sum _P \e _P A_{\a_{Q1},\cdots,\a_{QN}}(k_{PQ1},\cdots,k_{PQN})
\exp (i\sum ^N_{j=1} k_{P_j}x_j),
\eea
where the summation is taken over all permutations and negations of
$k_1,\cdots,k_N,$ and $Q$ is the permutation of the $N$ particles such that
$1\leq   x_{Q1}\leq   \cdots  \leq  x_{QN}\leq   L$.
The symbol $\e_P$ is a sign factor $\pm1$ and changes its sign
under each 'mutation'. Substituting the wavefunction into  the
eigenvalue equation $ H| \Psi  \rangle = E | \Psi \rangle $,
one gets
\bea
A_{\cdots,\a_j,\a_i,\cdots}(\cdots,k_j,k_i,\cdots)&=&S_{ij}(
    k_i,k_j)
    A_{\cdots,\a_i,\a_j,\cdots}(\cdots,k_i,k_j,\cdots),\no\\
A_{\a_i,\cdots}(-k_j,\cdots)&=&s^L(k_j;p_{1\a_i})A_{\a_i,\cdots}
    (k_j,\cdots),\no\\
A_{\cdots,\a_i}(\cdots,-k_j)&=&s^R(k_j;p_{L\a_i})A_{\cdots,\a_i}(\cdots,k_j),
\eea
where $S_{ij}(k_i,k_j) $ are
the two-particle scattering matrices,
\beq
S_{ij}(k_i,k_j)=\frac {\theta (k_i)-\theta (k_j)+ic P_{ij}}
{\theta (k_i)-\theta (k_j)+ic }\label{s-matrix}
\eeq
where $P_{ij}$ denotes the operator interchanging the species variables
$\a_i$ and $\a_j,(\a_i,\a_j=+,0,-)$,the rapidities $\theta(k_j)$ are
related to the single-particle quasi-momenta $k_j$ by $\theta (k)=\frac
{1}{2} \tan (\frac {k}{2})$ and the dependence on the system parameter
$g$ is incorporated in the parameter $c=1/g$.
 $~s^L(k_j;p_{1\a_i})$ and  
 $~s^R(k_j;p_{L\a_i})$ are the boundary scattering matrices,
\bea
s^L(k_j;p_{1\a_i})&=&\frac {1-p_{1\a_i}e^{ik_j}}
{1-p_{1\a_i}e^{-ik_j}},\no\\
s^R(k_j;p_{L\a_i})&=&\frac {1-p_{L\a_i}e^{-ik_j}}
{1-p_{L\a_i}e^{ik_j}}e^{2ik_j(L+1)}
\eea
with $p_{1\a_i}$ and $p_{L\a_i}$ being given by the following formulae,
corresponding to (\ref{boundary11} -- {\ref{boundary33}), respectively,
\bea
{\rm Case~(i):}~~
&&p_{1+}=p_{10}=p_{1-}\equiv p_1=
   -1-\frac {2g}{2-\xi^I_-},\no\\
&&p_{L+}=p_{L0}=p_{L-}\equiv p_L=
  -1- \frac {2g}{2-\xi^I_+};\label {p1}\\
{\rm Case~(ii):}~~
&&p_{1+}=-1,~~~p_{10}=p_{1-}=
   -1+\frac {2g}{\xi^{II}_-},\no\\
&&p_{L+}=-1,~~~p_{L0}=p_{L-}=
  -1+ \frac {2g}{\xi^{II}_+};\label {p2}\\
{\rm Case~(iii):}~~
&&p_{1+}=p_{10}=-1,~~~p_{1-}=
   -1+\frac {2g}{\xi^{III}_-},\no\\
&&p_{L+}=p_{L0}=-1,~~~p_{L-}=
  -1+ \frac {2g}{\xi^{III}_+}.\label {p3}
\eea

As is seen above, the two-particle S-matrix (\ref{s-matrix}) is 
nothing but the R-matrix of the $gl(3)$-invariant Heisenberg isotropic
magnetic chain
and thus satisfies the quantum Yang-Baxter equation (QYBE),
\beq
S_{ij}(k_i,k_j)S_{il}(k_i,k_l)S_{jl}(k_j,k_l)=
S_{jl}(k_j,k_l)S_{il}(k_i,k_l)S_{ij}(k_i,k_j).
\eeq
It can be checked that the boundary scattering matrices $s^L$ and $s^R$ 
obey the reflection equations:
\bea
&&S_{ji}(-k_j,-k_i)s^L(k_j;p_{1\a_j})S_{ij}(-k_i,k_j)s^L(k_i;p_{1\a
  _i})\no\\
&&~~~~~~~~~~~~~~~~~~=s^L(k_i;p_{1\a _i})S_{ji}(-k_j,k_i)s^L(k_j;p_{1\a _i})
  S_{ij}(k_i,k_j),\no\\
&&S_{ji}(-k_j,-k_i)s^R(k_j;p_{L\a_j})S_{ij}(k_i,-k_j)s^R(k_i;p_{L\a
  _i})\no\\
&&~~~~~~~~~~~~~~~~~~= s^R(k_i;p_{L\a _i})S_{ji}(k_j,-k_i)s^R(k_j;p_{L\a _i})
  ;p_{\a_i})S_{ji}(k_j,k_i).\label{reflection-e}
\eea
This is seen as follows. One introduces the notation
\beq
s(k;p)=\frac  {1-pe^{-ik}}{1-p e^{ik}}.
\eeq
Then the boundary scattering matrices $s^L(k_j;p_{1\a_i})$,
 $~s^R(k_j;p_{L\a_i})$ can be written as, corresponding to (\ref{p1}
-- {\ref{p3}), respectively,
\bea
{\rm Case~i:}~~&&s^L(k_j;p_{1\a_i})=s(-k_j;p_1)I,\no\\
&&s^R(k_j;p_{L\a_i})=e^{ik_j2(L+1)}s(k_j;p_L)I;\label{sa}\\
{\rm Case~ii:}~~&&s^L(k_j;p_{1\a_i})=s(-k_j;p_{1+})\lt(
\begin{array}{ccc}
1 &0 &0\\
0&\frac{\zeta_-+\l_j}
	       {\zeta_--\l_j} & 0\\
0&0&\frac{\zeta_-+\l_j}
	       {\zeta_--\l_j}
\end{array}
\rt),\no\\
&&s^R(k_j;p_{L\a_i})=e^{ik_j2(L+1)}s(k_j;p_{L+})\lt(
\begin{array}{ccc}
1&0&0\\
0&\frac{\zeta_+-\l_j}
	       {\zeta_++\l_j} & 0\\
0&0&\frac{\zeta_+-\l_j}
	       {\zeta_++\l_j}
\end{array}
\rt);\label{sb}\\
{\rm Case~iii:}~~&&s^L(k_j;p_{1\a_i})=
s(-k_j;p_{1+})\lt(
\begin{array}{ccc}
1 &0 &0\\
0&1   & 0\\
0&0&\frac{\k_-+\l_j}
	       {\k_--\l_j}
\end{array}
\rt),\no\\
&&s^R(k_j;p_{L\a_i})=e^{ik_j2(L+1)}s(k_j;p_{L+})\lt(
\begin{array}{ccc}
1&0&0\\
0&1&0\\
0&0&\frac{\k_+-\l_j}
	       {\k_++\l_j} 
\end{array}
\rt),\label{sc}
\eea
where $I$ stands for $3\times 3$ identity matrix and 
$p_{1+},~p_{L+}$ are the ones given in (\ref{p2}); $\zeta _{\pm},~
\k _{\pm}$ are parameters defined by
\beq
\zeta _{\pm}=\frac {i(g-\xi^{II}_{\pm})}{2g},~~~~~~
\k_{\pm}=\frac {i(g-\xi^{III}_{\pm})}{2g}.
\eeq
The boundary scattering matrices for cases (iv) -- (ix) can be easily
built from (\ref{sa}), (\ref{sb}) and (\ref{sc}).
We immediately see that (\ref{sa}) are the trivial solutions of the reflection
equations (\ref{reflection-e}), whereas (\ref{sb}) and (\ref{sc})
are the diagonal
solutions \cite{Skl88,Mez91}. 

The diagonalization of Hamiltonian (\ref{hamiltonian}) reduces 
to solving  the following matrix  eigenvalue equation
\beq
T_jt= t,~~~~~~~j=1,\cdots,N,
\eeq
where $t$ denotes an eigenvector on the space of the spin variables
and $T_j$ takes the form
\beq
T_j=S_j^-(k_j)s^L(-k_j;p_{1\s_j})R^-_j(k_j)R^+_j(k_j)
    s^R(k_j;p_{L\s_j})S^+_j(k_j)
\eeq
with
\bea
S_j^+(k_j)&=&S_{j,N}(k_j,k_N) \cdots S_{j,j+1}(k_j,k_{j+1}),\no\\
S^-_j(k_j)&=&S_{j,j-1}(k_j,k_{j-1})\cdots S_{j,1}(k_j,k_1),\no\\
R^-_j(k_j)&=&S_{1,j}(k_1,-k_j)\cdots S_{j-1,j}(k_{j-1},-k_j),\no\\
R^+_j(k_j)&=&S_{j+1,j}(k_{j+1},-k_j)\cdots S_{N,j}(k_N,-k_j).
\eea
This problem can  be solved using the algebraic Bethe ansatz method.
The Bethe ansatz equations are 
\bea
e^{ik_j2(L+1)}F(k_j;p_{1+},p_{L+})
&=&\prod_{\s=1}^{M_1}\frac{\t_j-\l^{(1)}_\s+ic/2}
      {\t_j-\l^{(1)}_\s-ic/2}
\frac{\t_j+\l^{(1)}_\s+ic/2}
      {\t_j+\l^{(1)}_\s-ic/2},\no\\
\prod_{j=1}^N\frac{\l^{(1)}_\s-\t_j+ic/2}{\l^{(1)}_\s-\t_j-ic/2}
\frac{\l^{(1)}_\s+\t_j+ic/2}{\l^{(1)}_\s+\t_j-ic/2}&=&
   G(\l ^{(1)}_\s;\zeta _-,\zeta_+) \prod_{\stackrel {\rho=1}
   {\rho \neq \s}}^{M_1}\frac{\l^{(1)}_\s-\l^{(1)}_\rho+ic}
   {\l^{(1)}_\s-\l^{(1)}_\rho-ic}
   \frac{\l^{(1)}_\s+\l^{(1)}_\rho+ic}
   {\l^{(1)}_\s+\l^{(1)}_\rho-ic}\no\\
& &   \prod_{\rho=1}^{M_2}\frac{\l^{(1)}_\s-\l^{(2)}_\rho-ic/2}
   {\l^{(1)}_\s-\l^{(2)}_\rho+ic/2}
   \frac{\l^{(1)}_\s+\l^{(2)}_\rho-ic/2}
   {\l^{(1)}_\s+\l^{(2)}_\rho+ic/2},\no\\
   & &\s=1,\cdots,M_1,\no\\
\prod_{\rho=1}^{M_1}\frac{\l^{(2)}_\g-\l^{(1)}_\rho+ic/2}
   {\l^{(2)}_\g-\l^{(1)}_\rho-ic/2}
\frac{\l^{(2)}_\g+\l^{(1)}_\rho+ic/2}
   {\l^{(2)}_\g+\l^{(1)}_\rho-ic/2}&=&
   K(\l^{(2)}_\g;\k_-,\k_+)\prod_{\stackrel {\rho=1}{\rho \neq \g}}^{M_2}
   \frac{\l^{(2)}_\g-\l^{(2)}_\rho+ic}
   {\l^{(2)}_\g-\l^{(2)}_\rho-ic}
   \frac{\l^{(2)}_\g+\l^{(2)}_\rho+ic}
   {\l^{(2)}_\g+\l^{(2)}_\rho-ic},\no\\
  & & \g=1,\cdots,M_2,\label{Bethe-ansatz}
\eea
where 
\bea
F(k_j;p_{1+},p_{L+})&=& s(k_j;p_{1+}) s(k_j;p_{L+}),({\rm for\; all\; cases})      \no\\
G(\l^{(1)}_\s;\zeta _-,\zeta _+)&=& \left \{ \begin {array}{ll}
1 & case \;(i)\\
\frac {(\zeta _-+\lambda ^{(1)}_\s +\frac {ic}{2})}
 {(\zeta _--\lambda ^{(1)}_\s +\frac {ic}{2})}
\frac {(\zeta _++\lambda ^{(1)}_\s +\frac {ic}{2})}
 {(\zeta _+-\lambda ^{(1)}_\s +\frac {ic}{2})}
& case\;( ii)\\
1 & case \;(iii)\\
\frac {(\zeta _++\lambda ^{(1)}_\s +\frac {ic}{2})}
 {(\zeta _+-\lambda ^{(1)}_\s +\frac {ic}{2})}
& case \;(iv)\\
\frac {(\zeta _-+\lambda ^{(1)}_\s +\frac {ic}{2})}
 {(\zeta _--\lambda ^{(1)}_\s +\frac {ic}{2})}
& case \;(v)\\
1 & case \;(vi)\\
1 & case \;(vii)\\
\frac {(\zeta _-+\lambda ^{(1)}_\s +\frac {ic}{2})}
 {(\zeta _--\lambda ^{(1)}_\s +\frac {ic}{2})}
& case \;(viii)\\
\frac {(\zeta _++\lambda ^{(1)}_\s +\frac {ic}{2})}
 {(\zeta _+-\lambda ^{(1)}_\s +\frac {ic}{2})}
& case\;( ix)
\end {array} \right. .\no\\
K(\l^{(2)}_\g;\k_-,\k_+)&=& \left \{ \begin {array}{ll}
1 & case\;( i)\\
1 & case \;(ii)\\
\frac {(\k _-+\lambda ^{(2)}_\g +ic)}
 {(\k _--\lambda ^{(2)}_\g +ic)}
\frac {(\k _++\lambda ^{(2)}_\g +ic)}
 {(\k _+-\lambda ^{(2)}_\g +ic)}
& case \;(iii)\\
1 & case \;(iv)\\
1 & case \;(v)\\
\frac {(\k _++\lambda ^{(2)}_\g +ic)}
 {(\k _+-\lambda ^{(2)}_\g +ic)}
& case \;(vi)\\
\frac {(\k _-+\lambda ^{(2)}_\g +ic)}
 {(\k _--\lambda ^{(2)}_\g +ic)}
& case \;(vii)\\
\frac {(\k _++\lambda ^{(2)}_\g +ic)}
 {(\k _+-\lambda ^{(2)}_\g +ic)}
& case \;(viii)\\
\frac {(\k _-+\lambda ^{(2)}_\g +ic)}
 {(\k _--\lambda ^{(2)}_\g +ic)}
& case \;(ix)
\end {array} \right. .\no\\
\eea
The energy eigenvalue $E$ of the model is given by
$E=-2\sum ^N_{j=1}\cos k_j$ (modular an unimportant additive constant coming
from the chemical potential term).

\sect{Conclusion}

In conclusion, we have studied integrable open-boundary conditions for the
eight-state supersymmetric $U$ model. The  quantum integrability of the
system follows from the fact
that the Hamiltonian may be embedded into
a one-parameter family of commuting transfer matrices. Moreover, the Bethe
ansatz equations are derived by use of the coordinate space Bethe ansatz
approach. This provides us with a basis for computing the finite size
corrections (see, e.g. \cite{Asa96})
to the low-lying energies in the system, which in turn allow
us to use the boundary conformal field theory technique to study
the critical properties of the boundary.
The details will be treated in a separate publication.

\vskip.3in
\acknowledgments
Y.-Z.Z was supported by the Queen Elizabeth II Fellowship Grant from 
Australian Research Council.


\end{document}